\documentclass[11pt]{article}
\usepackage{mathrsfs} 
\usepackage{mathpazo}
\usepackage{fullpage}
\usepackage{times}
\usepackage{hyperref}
\usepackage{graphicx}

\usepackage{amsmath}
\usepackage{amssymb}
\usepackage{amsfonts}
\usepackage{dsfont}
\usepackage{url}
\usepackage[final]{changes} 
\usepackage[applemac]{inputenc}
\usepackage[affil-it]{authblk} 

\newcommand{\beq}{\begin{equation}}
\newcommand{\eeq}{\end{equation}}
\newcommand{\bea}{\begin{eqnarray}}
\newcommand{\eea}{\end{eqnarray}}

\def\be{\begin{equation}}
\def\ee{\end{equation}}
\def\bea{\begin{eqnarray}}
\def\eea{\end{eqnarray}}

\newcommand{\pup}[1]{\ensuremath{P_{U}}}

\newcommand{\hgate}{{\sf H}}
\newcommand{\pgate}{{\sf P}}
\newcommand{\rgate}{{\sf R}}
\newcommand{\xgate}{{\sf X}}

\newcommand{\zgate}{{\sf Z}}
\newcommand{\cnot}{{\sf CNOT}}

\newcommand{\hreg}[1]{\ensuremath{\mathcal{#1}}}

\def\diam#1{\|#1\|_\diamond}
\def\trnorm#1{\|#1\|_1}

\newcommand{\simul}[1]{\ensuremath{\mathscr{S}_{#1}}}

\newcommand{\rhoin}{\ensuremath{\rho_{\mathrm{in}}}}
\newcommand{\dens}[1]{\ensuremath{\mathrm{D}(#1)}}

\newcommand{\linop}[1]{\ensuremath{\mathrm{L}(#1)}}
\newcommand{\lin}[2]{\ensuremath{\linop{#1,#2}}}

\input{qcircuit} 
\newcommand{\dw}[1][-1]{\ar @{--} [0,#1]}

\newcommand{\twowire}[1]{\ar@{}[d]_-{\txt{#1}\Bigg\{}}

\begin{document}
\title{Delegating Private Quantum Computations}

\author{Anne Broadbent}
\affil{Department of Mathematics and Statistics\\ University of Ottawa\\ \url{abroadbe@uottawa.ca}}

\date{}

\normalem
\maketitle
\begin{abstract}
We give a protocol for the delegation of quantum computation on
encrypted data. More specifically, we show that in a client-server
scenario, where the client holds the encryption key for an encrypted
quantum register held by the server, it is possible for the server
to perform a universal set of quantum gates on the quantum data. All
Clifford group gates are non-interactive, while the remaining
non-Clifford group gate that we implement (the $\pi/8$ gate)
requires the client to prepare and send a single random auxiliary
qubit (chosen among four possibilities), and exchange classical
communication. This construction improves on previous work, which
requires either multiple auxiliary qubits or two-way quantum
communication. Using a reduction to an entanglement-based protocol,
we show privacy against any adversarial server according to a
simulation-based security definition.
\end{abstract}

\maketitle

\section{Introduction}

Today's computing paradigm displays seemingly contradictory
requirements. On one hand, computations are often delegated to
remote powerful computing centers, while on the other hand, the data
that is being processed is expected to remain private. We thus face
the conundrum of wanting to compute on encrypted data. One specific
scenario that allows data to be encrypted by one party and processed
by another is known as \emph{fully homomorphic
encryption}~\cite{G09, RAD78}.

This paper\footnote{This paper was initially submitted in its current form (up to minor corrections, updates and formatting) to the Proceedings of the 32nd International Cryptology Conference (CRYPTO 2012), where it was rejected. The results were then improved to include an experimental demonstration and eventually appeared as~\cite{QCED}. Thus this version appears here for the first time in print and will be of special interest to those wishing to focus on the theory contribution in~\cite{QCED}.}
 addresses the problem of performing \emph{quantum}
computations on encrypted \emph{quantum} data. In one way, we relax
the requirements of fully homomorphic encryption by allowing
\emph{interaction}, but at the same time, we strengthen the
requirements by asking for information-theoretic security. This is
an asymmetric scenario---it deals with a quantum server (or quantum
\emph{cloud} architecture), a
 particularly relevant scenario  due to the current challenges
in building quantum computational devices. This scenario is also
considered in~\cite{ABE10, BFK09, C05}. We show that an
almost-classical client can delegate the execution of any quantum
computation  to a remote quantum server, and that this computation
can be performed on quantum data that is encrypted via the quantum
one-time pad~\cite{AMTW00}; informally, privacy is maintained since
the server \emph{never} learns the encryption key. An important
requirement of any  protocol for delegated computation on encrypted
data is that the operations performed by the client should be
significantly easier to perform than the computation itself. In our
scenario, we achieve this since  the client does not require the
capacity of universal quantum computation. She only requires the
ability to perform encryption and decryption (for this she needs to
be able to apply single-qubit Pauli operators); she also must be
able to prepare send random qubits chosen from a set of four
possibilities. This set of states is unitarily equivalent to the set
$\{\ket{0}, \ket{1}, \ket{+}, \ket{-}\}$, which are known as the
BB84 states, for the crucial role that they play in the quantum key
distribution protocol known by the same name~\cite{BB84}.
Such a client does not require quantum memory and can be implemented
with current technology, for instance, using photon
polarization~\cite{springerlink:10.1007/BF00191318,RevModPhys.79.135}.
We suppose that the client is honest and prove security against any
cheating server via simulations. Similar functionality has been
achieved before~\cite{ABE10,  BFK09, C05}.

  Compared to prior work, our contribution has the advantage of providing a
 conceptually simple proof of correctness, together with a security
 definition and proof that is applicable to all types of prior
 information, including shared entanglement. Additionally, our
protocol is more efficient in terms of quantum and classical
 communication. Compared to~\cite{BFK09}, our gain for a general quantum computation
 is by a constant factor; nevertheless, this means that, using current
 technology~\cite{BKBFZW12},  our protocol could lead to the experimental
 delegation of a wider class of private quantum computations.

A sample application of our protocol would be the delegated, private
execution of Shor's algorithm~\cite{Shor97} which can be used to
factor in polynomial time on a quantum computer (this computation is
widely believed to be intractable on a classical computer). Since
the computation is performed on an encrypted input, the server will
not know which integer he is factoring; if this integer corresponds
to an RSA public key~\cite{RSA78} then the server will not know
which public key he is helping to break.
 We thus see that
quantum computing on encrypted data is useful for the delegation of
problems that can be solved in quantum polynomial time, with the
underlying assumption that they cannot be solved in classical
polynomial time.

However, applications of delegating private quantum computations are
foreseeable \emph{even if} it turns out that quantum computers are
no more powerful than classical ones, since delegated computation on
encrypted \emph{quantum} data is also achieved. This could be
useful, for instance, to enable a client (with no universal quantum
computer) to perform quantum circuits on quantum data such as
\emph{quantum money}~\cite{FGHLS10} or \emph{quantum
coins}~\cite{MS10}.

\section{Contributions and Related Work}
\label{sec:contributions}

In order to achieve our results, we consider the scenario of quantum
computing on encrypted data: one party holds a quantum register
while the other party holds the encryption key. It is well known
that performing a Clifford group circuit (or, more generally, a
stabilizer circuit) on quantum data encrypted with the one-time pad
can be achieved non-interactively: the server (holding the encrypted
data) applies the target gates, while the client (holding secret
encryption keys) simply adjusts her knowledge of the encryption key
(see Sections~\ref{sec:measure-prepare} and~\ref{sec:Clifford}).
What remains in order to perform a universal quantum computation is
to show how to implement a non-Clifford group gate on encrypted
data.

Our main contribution is a simple protocol for computing a $\pi/8$
gate over encrypted data (Figure~\ref{fig:R-gate}). We define
security via simulation and show that the final protocol is secure
against any malicious server (Section~\ref{sec:correct-secure}). We
use as proof technique the method of transforming a qubit-based
protocol into an equivalent protocol that is more easily  proved
secure, but that involves entanglement. This technique is attributed
to Shor and Preskill~\cite{SP00}, who used it in the context of
proving the security of the BB84~\cite{BB84} quantum key exchange
protocol, and has since appeared in the context of quantum message
authentication~\cite{BCGST02} and cryptography in the
bounded-quantum-storage model~\cite{DFSS05}.

 We emphasize that the protocol achieves the same level of privacy
as the quantum one-time pad, which is the highest possible level of
security:  it depends only on the correctness of quantum mechanics
and in particular does not rely on any computational assumptions. In
contrast, fully homomorphic encryption~\cite{G09} provides
computational security only because it uses a public-key encryption
scheme.


 We have phrased our contribution in terms of performing a
\emph{publicly-known} circuit on encrypted data. Hiding the entire
computation is possible simply by executing a universal circuit on
an encrypted input, part of which contains the description of the
target circuit to be implemented. Furthermore, the protocol can
easily be adapted to allow the server to provide an input.

Previous results achieve similar (or even identical) functionality,
with an similar level of security, but require more resources:
\begin{enumerate}
\item The \emph{secure assisted quantum computation} protocol of
Childs~\cite{C05}, accomplishes the same functionality as our
protocol, but with a significant difference in that the protocol
involves two-way quantum communication and the client needs to be
able to execute a two-qubit \emph{swap} gate. We give more details
on how our protocol differs from Childs' in
Section~\ref{sec:non-clifford}.

\item The protocol for \emph{universal blind quantum computation} of
Broadbent, Kashefi and Fitzsimmons~\cite{BFK09} achieves a similar
functionality, by using more resources. While the goal of blind
quantum computing is first and foremost to hide the computation
itself, the protocol also achieves computation on encrypted data.
Because our protocol does not require that the circuit be hidden, we
manage to reduce the requirements in terms of communication:
while~\cite{BFK09} requires for each gate (including the identity),
24 bits of forward communication, eight bits of backward
communication and eight auxiliary qubits, our protocol reduces the
communication to null for all but the execution of a non-Clifford
group gate; in the specific case of the $\pi/8$ gate, the
interaction consists of a single auxiliary qubit and two classical
bits (one bit in each direction). Furthermore, \cite{BFK09} requires
that auxiliary qubits be prepared from a set of eight possible
states, while we manage to reduce this to~four (see
Section~\ref{sec:non-clifford}). Note that the universal blind
quantum computation protocol was recently experimentally
demonstrated~\cite{BKBFZW12}.

\item The protocol for \emph{quantum prover interactive
proofs} of Aharonov, Ben-Or and Eban \cite{ABE10} establishes, on
top of the functionality that we implement, a \emph{verification}
mechanism to ensure that the server is performing the correct
computation. The cost of this construction is that the client needs
to prepare auxiliary quantum systems of size polynomial in the
parameter determining the security of the verification. Our protocol
does not provide any verification mechanism, but manages to
significantly limit the quantum power needed by the client.
\end{enumerate}

Finally, our work is related to the more general scenario of
two-party quantum computation, where it is the case that \emph{both}
parties hold keys to the encrypted quantum data. Dupuis, Nielsen and
Salvail~\cite{DNS10} gave a protocol for two-party secure
quantum computation in the case of \emph{specious} (a version of
quantum semi-honest) adversaries. Certain of our sub-protocols
display a similarity (local Clifford group gates are essentially
identical). However, our contribution for the $\rgate$-gate protocol
is very different since it requires no quantum interaction. In this
respect, the work of \cite{DNS10} is closer to the work
of~\cite{C05}.\looseness=-1

\section{Preliminaries}
\label{sec:preliminaries} We assume the reader is familiar with the
basics of quantum information~\cite{NC00}. Recall the following
notation: $\ket{+} = \frac{1}{\sqrt{2}} (\ket{0} + \ket{1}$), the
Pauli gates  $\xgate: \ket{j}
  \mapsto \ket{j \oplus 1}$ and $\zgate: \ket{j} \mapsto (-1)^j\ket{j}$, as well as
  the single-qubit
  Hadamard and phase gates, $\hgate : \ket{j} \mapsto
  \frac{1}{\sqrt{2}} (\ket{0} + (-1)^j \ket{1}$), $\pgate: \ket{j} \mapsto (i)^j\ket{j}$. 
Recall also the two-qubit gate  $\cnot: \ket{j}\ket{k} \mapsto
\ket{j}\ket{j \oplus k}$. An \emph{EPR-pair} is a pair of maximally
entangled qubits, $\frac{1}{\sqrt{2}}(\ket{00} + \ket{11})$.

\subsection{Quantum circuits and circuit identities}
\label{sec:prelim-circuits} The \emph{Clifford Group}~\cite{Gott98}
is the set of operators that conjugate Pauli operators into Pauli
operators.  A universal gate set for \emph{Clifford group circuits}
consists of the Pauli gates themselves, together with \hgate,
$\pgate$ and \cnot. \emph{Stabilizer circuits} are formed by adding
the operations of single-qubit measurements and auxiliary qubit
preparation to the Clifford group circuits. Stabilizer circuits are
not universal for quantum computation~\cite{Gott98}, however
supplementing with \emph{any} additional gate outside of the group
(such as $\rgate: \ket{j} \mapsto e^{ij\pi/4}\ket{j}$ or the Toffoli
gate $\textsf{T}: \ket{j}\ket{k}\ket{\ell} \mapsto
\ket{j}\ket{k}\ket{\ell \oplus jk}$) is necessary and sufficient for
universality. We interchangeably refer to the \rgate-gate as the
$\pi/8$~gate.\looseness=-1

We will make use of the following identities which all hold up to an
irrelevant global phase: $ \xgate \zgate = \zgate \xgate$, $\pgate
\zgate = \zgate \pgate$, $\pgate \xgate = \xgate \zgate \pgate$,
$\rgate \zgate = \zgate \rgate$, $\rgate \xgate = \xgate \zgate
\pgate \rgate$,  $\pgate^2  = \zgate$ and $\pgate^{a\oplus b} =
\zgate^{a\cdot b} \pgate^{a + b}$ (for $a, b \in \{0,1\}$).

\begin{figure}[h!]
\label{circuit-proof:X-2}
 \centerline{
 \Qcircuit @C=1em @R=1em {
\lstick{\ket{\psi}} & \qw &\targ      &  \meter & \cw   &  \rstick{c}      \\
\lstick{\ket{+}}    & \qw &\ctrl{-1}  &   \qw  & \qw &
\rstick{\xgate^c\ket{\psi}}&   }
 }
 \caption{$\xgate$-teleportation circuit~\cite{ZLC00}.
 \label{fig:X-teleport}}
\end{figure}
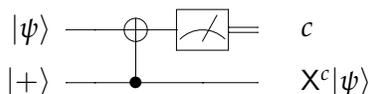

In order to derive and prove our results, we make use of known
techniques for manipulating quantum circuits. Of significant
relevance to our work are the techniques developed by Childs, Leung,
and Nielsen~\cite{CLN05} to manipulate circuits that produce an
output that is correct \emph{up to known Pauli corrections}. These
techniques are based on a variant of teleportation introduced by
Zhou, Leung, and Chuang~\cite{ZLC00} (see
Figure~\ref{fig:X-teleport}, as well as
Appendix~\ref{sec:appendix-R-gate} for a derivation of this circuit
identity). Here and in the following figures, measurements are
performed in the computational basis.

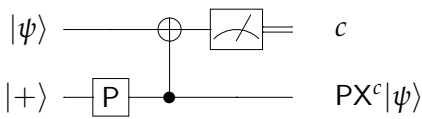
\begin{figure}[h!]
 \centerline{
 \Qcircuit @C=1em @R=1em {
\lstick{\ket{\psi}} & \qw &\targ      &  \meter & \cw   &  \rstick{c}      \\
\lstick{\ket{+}}    & \gate{\pgate} &\ctrl{-1}  &   \qw  & \qw &
\rstick{\pgate\xgate^c\ket{\psi}}&   }
 }
 \caption{Circuit identity: the $\pgate$-gate commutes with control. A similar identity holds if we replace the $\pgate$-gate with a $\zgate$-gate.}
 \label{fig:P-commutes}
\end{figure}

\noindent We also use the fact that $\pgate$ and $\zgate$ commute
with control (Figure~\ref{fig:P-commutes}).

\begin{figure}[h!]
 \centerline{
 \Qcircuit @C=1em @R=1em {
&\lstick{\ket{0}} & \gate{\hgate} &\ctrl{1}      & \qw & \qw & \qw& \rstick{\zgate^d\pgate^y\ket{+}}      \\
&\lstick{\ket{0}}    & \qw &\targ  &   \gate{\pgate^y}  &
\gate{\hgate} &  \meter & \cw   &  \rstick{d}
  \gategroup{1}{1}{2}{4}{1.4em}{--}
}
 }
 \caption{Circuit identity:  entanglement-based circuit that prepares a
qubit~$\zgate^d\pgate^y\ket{+}$ for uniformly random bits $y$
and~$d$ (here, $y$ is chosen uniformly at random, and $d$ is
determined by the measurement). The circuit in the dashed box
prepares an EPR-pair. }
 \label{fig:qubit-prep}
\end{figure}
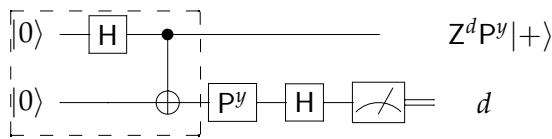

\noindent Finally, we make use of an entanglement-based circuit that
prepares a qubit~$\zgate^d\pgate^y\ket{+}$ for uniformly random bits
$y$ and $d$ (Figure~\ref{fig:qubit-prep}). Correctness of this
circuit is easy to verify.

\subsection{Classical and quantum encryption}
\label{sec:encryption}
 The classical \emph{one-time pad} is an
encryption procedure that maps each bit $j$ of a plaintext to $j
\oplus r$ for a uniformly random key bit~$r$ (which we denote $r
\in_R \{0,1\}$). Since the ciphertext $j \oplus r$ is uniformly
random (as long as $r$ is unknown), the plaintext~$j$ is perfectly
concealed. The \emph{quantum one-time pad}~\cite{AMTW00} is the
quantum analog of the classical one-time pad. The encryption
procedure for a single qubit consists of uniformly randomly applying
an operator in~$\{\textsf{I}, \xgate, \zgate, \xgate\zgate \}$, or
equivalently, applying $\xgate^a\zgate^b$ for uniformly random bits
$a$ and~$b$ (here, $\textsf{I}$ is the identity). This maps any
single qubit to the maximally mixed state on one qubit, which we
denote~$\mathds{1}_2$; thus the quantum plaintext is perfectly
concealed.

\subsection{Quantum registers and channels}
A \emph{quantum register} is a collection of qubits in some finite
dimensional Hilbert space, say $\hreg{X}$. We denote
$\dens{\hreg{X}}$ the set of density operators acting on $\hreg{X}$.
The set of all linear mappings from \hreg{X}\ to \hreg{Y}\ is
denoted by \lin{\hreg{X}}{\hreg{Y}}, with  \linop{\hreg{X}} being a
shorthand for \lin{\hreg{X}}{\hreg{X}}. A linear super-operator
$\Phi : \linop{\hreg{X}} \rightarrow \linop{\hreg{Y}}$ is
\emph{admissible} if it is completely positive and trace-preserving.
Admissible super-operators represent mappings from density operators
to density operators, that is, they represent the most general
quantum maps.

Given admissible super-operators $\Phi$ and $\Psi$ that agree on
input space $\linop{\hreg{X}}$ and output space $\linop{\hreg{Y}}$,
we are interested (for cryptographic purposes) in characterizing how
``indistinguishable'' these processes are. The \emph{diamond norm}
provides such a measure: given that $\Phi$ or~$\Psi$ is applied with
equal probability, the optimal procedure to determine the identity
of the channel with only one use succeeds with probability $1/2 +
\diam{\Phi - \Psi}/4$. Here,
$\diam{\Phi - \Psi}= \max \{ \trnorm{(\Phi \otimes
\mathds{1}_{\hreg{W}} )(\rho) - (\Psi\otimes \mathds{1}_{\hreg{W}}
)(\rho)} : \rho \in \dens{\hreg{X} \otimes \hreg{W}} \}\,,$
where $\hreg{W}$ is any space with dimension equal to that of
$\hreg{X}$ and $\mathds{1}_{\hreg{W}}$ is the identity
in~\linop{\hreg{W}}, and where the \emph{trace norm} of an operator
$X$ is defined as $\trnorm{X} = \text{Tr}\sqrt{X^*X}$\,.

\section{Delegating private quantum computations}
\label{sec:delegate-priv-comp}
 A general quantum circuit can be
decomposed into a sequence of the following: gates in
$\{\xgate, \zgate, \hgate, \pgate, \cnot, \rgate\}$, auxiliary qubit
preparation in~$\ket{0}$ and single-qubit computational basis
measurements (strictly speaking, this set is redundant; the choice
of these gates will become clear later). We show in the following
sections that these operations can be executed by a server who has
access only to the input in its encrypted form (where the encryption
is the quantum one-time pad), and we show that the output can
nevertheless be decrypted by the client, and that the server does
not learn anything about the input. In order to accomplish this, we
give a series of protocols, each accomplishing the execution of a
circuit element. For each such protocol, the client (who knows the
encryption key for the input to the protocol) can compute a
decryption key that, if applied to the output of the protocol, would
result in the output of the circuit element applied to the
unencrypted input.

Sections~\ref{sec:measure-prepare} and~\ref{sec:Clifford} show
protocols for stabilizer circuit elements, while
Section~\ref{sec:non-clifford} gives a protocol for a non-Clifford
group gate (this is the only gate that uses interaction). In all
cases, we give explicit constructions for pure states; it is
straightforward to verify that the same constructions work on
systems that are entangled.


We can see each of the protocols as gadgets that implement a circuit
element, up to a known re-interpretation of the key. In order to
execute a larger target circuit, each of its circuit elements is
executed in sequence as one of these gadgets; it is sufficient for
the client to re-adjust her knowledge of the encryption keys on each
relevant quantum wire after each gadget. Our  protocol for quantum
computing on encrypted data is given as:

\begin{enumerate}
 \item \label{prot:1}The client encrypts her register with the quantum one-time pad and sends the encrypted register to the server.
 \item  \label{prot:2}The client and server perform the gadgets as given in Section~\ref{sec:measure-prepare}--\ref{sec:non-clifford},
  according to the circuit that is to be executed, with the client re-adjusting the encryption keys on each relevant
quantum wire  after each gadget.
 \item  \label{prot:3}The server returns the output register to the client, who decrypts it according to the key that she has computed.
\end{enumerate}

Note that this high-level protocol does not involve any interaction
other than the sending (Step~\ref{prot:1}) and receiving
(Step~\ref{prot:3}) of the encrypted data. Only a \emph{single}
gadget in the implementation of Step~\ref{prot:2} in our
construction is interactive (Section~\ref{sec:non-clifford}), and
the quantum part of this interaction is one-way (from the client to
the server), consisting of the sending of a single  random auxiliary
qubit in $\{\frac{1}{\sqrt{2}}(\ket{0} + \ket{1}),
\frac{1}{\sqrt{2}}(\ket{0} + i\ket{1}), \frac{1}{\sqrt{2}}(\ket{0} -
\ket{1}), \frac{1}{\sqrt{2}}(\ket{0} - i\ket{1})\}$,
which can be sent, without loss of generality, at the beginning of
the protocol.  Thus our protocol for quantum computing on encrypted
data is interactive, but completely classical, except for the
initial sending of auxiliary random single-qubit states, as well as
for the sending of the input and output registers (if they are
quantum). Furthermore, a target circuit that implements only
Clifford group operations can be executed with no interaction at all
(except for the interaction in Steps~\ref{prot:1} and~\ref{prot:3}).
We now proceed to give the protocols
(Sections~\ref{sec:measure-prepare}, \ref{sec:Clifford} and
\ref{sec:non-clifford}). Security is formally defined via simulation
and proved in Section~\ref{sec:correct-secure}.

\subsection{Protocols for measurement and auxiliary qubit
preparation} \label{sec:measure-prepare}

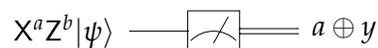
\begin{figure}[ht]
\centerline{ \Qcircuit @C=2em @R=1.5em {
   \lstick{\xgate^a \zgate^b \ket{\psi }}   &   \meter    &      \rstick{a \oplus y} \cw
} } \caption{\label{fig:measurement} Protocol for measurement.
Here,~$y$ denotes the outcome of the measurement on the unencrypted
input~$\ket{\psi}$.}
\end{figure}

The protocol for measuring a single qubit in the computational basis
is given in Figure~\ref{fig:measurement}: the server simply performs
a measurement on the encrypted qubit. The corresponding wire thus
goes from a quantum wire (with encryption operation
$\xgate^a\zgate^b$) to a classical wire (with  encryption key $a$);
the client can easily take this into account.

\begin{figure}[ht]
\centerline{
\Qcircuit @C=2em @R=1.5em {
   \lstick{ \ket{0 }}   &   \qw    &      \rstick{\xgate^0 \zgate^0\ket{0}} \qw
} } \caption{\label{fig:aux-prep} Protocol for auxiliary qubit
preparation.}
\end{figure}
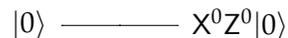

\noindent As represented in Figure~\ref{fig:aux-prep}, the server
may prepare an unencrypted auxiliary qubit in the $\ket{0}$ state
and incorporate it into the computation. The client simply sets the
encryption key for this qubit to be~0.\looseness=-1

\subsection{Protocols for Clifford group gates} \label{sec:Clifford}

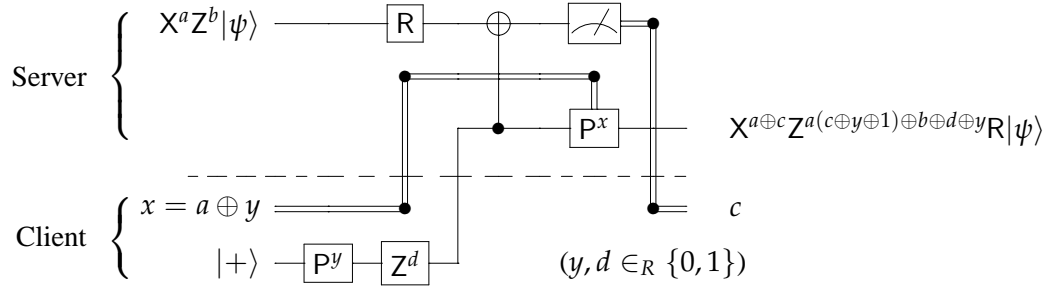
\begin{figure}[h!t]
\centerline{
\Qcircuit @C=2em @R=1.5em {
   \lstick{\xgate^a \zgate^b\ket{\psi}}   &   \gate{\xgate} &       \rstick{\xgate^a \zgate^b \xgate \ket{\psi }} \qw
} } \caption{\label{fig:X-gate} Protocol for an \xgate-gate.}
\end{figure}

 A series of well-known relationships between
the Pauli matrices and Clifford group operations~\cite{Gott98} is
the basis for the protocols given in
Figures~\ref{fig:X-gate}--\ref{fig:CNOT-gate}. Specifically, since
the $\xgate$ and $\zgate$ gates commute or anti-commute (and since
we can safely ignore a global phase), Figures~\ref{fig:X-gate} and
\ref{fig:Z-gate} can easily be seen as implementing a protocol for
the $\xgate$ and $\zgate$-gates. Similarly, the relation
$\hgate\xgate = \zgate\hgate$ is sufficient to show the
$\hgate$-gate protocol given in Figure~\ref{fig:H-gate}, and the
facts that $\pgate\zgate=\zgate\pgate$ and $\pgate\xgate =
-i\zgate\xgate\pgate$ show the $\pgate$-gate protocol given in
Figure~\ref{fig:P-gate}. Finally, the protocol for the \cnot-gate as
given in Figure~\ref{fig:CNOT-gate} can be verified in a similar
way.

Strictly speaking, in order to achieve universality, we do not need
all of the protocols given above: once we have the protocol for the
\rgate-gate (given in Section~\ref{sec:non-clifford} below), it is
sufficient to combine it with the protocols for \cnot\ and \hgate\
for universality. This can be seen since $\pgate=\rgate^2$,
$\zgate=\pgate^2$ and $\xgate=\hgate\zgate\hgate$. However, each of
these decompositions requires at least two \rgate-gates, and as we
will see below, the protocol for an \rgate-gate is relatively
expensive (it uses an auxiliary qubit and classical interaction). It
can thus be preferable to decompose a circuit into the redundant
gate set that we have used as this reduces the cost of many gates.
Also, by giving explicit protocols for all of these circuit
elements, we have established that a stabilizer circuit can be
performed on encrypted data \emph{without any} interaction
whatsoever, except for the exchanging of the encrypted data
register.


\begin{figure}[h!t]
\centerline{
\Qcircuit @C=2em @R=1.5em {
   \lstick{\xgate^a \zgate^b\ket{\psi}}   &   \gate{\zgate}    &       \rstick{\xgate^a \zgate^b \zgate \ket{\psi }} \qw
} } \caption{\label{fig:Z-gate} Protocol for a \zgate-gate.}
\end{figure}

\begin{figure}[h!t]
\centerline{
\Qcircuit @C=2em @R=1.5em {
   \lstick{\xgate^a \zgate^b\ket{\psi}}   &   \gate{\hgate}    &       \rstick{\xgate^b \zgate^a \hgate \ket{\psi }} \qw
} } \caption{\label{fig:H-gate} Protocol for an \hgate-gate.}
\end{figure}

\begin{figure}[h!t]
\centerline{
\Qcircuit @C=2em @R=1.5em {
   \lstick{\xgate^a \zgate^b\ket{\psi}}   &   \gate{\pgate}    &       \rstick{\xgate^a  \zgate^{a + b} \pgate \ket{\psi }} \qw
} } \caption{\label{fig:P-gate} Protocol for a \pgate-gate.}
\end{figure}

\begin{figure}[h!t]
\hspace{-0.75cm}
\centerline{
\Qcircuit @C=1em @R=1.5em {
                                   &  \ctrl{1}    &    \qw  \\ 
   \lstick{\hspace{-1cm}\raisebox{1.6em}{$(\xgate^a \zgate^b \otimes \xgate^c \zgate^d)\ket{\psi}\,\,\,\,\,$}} &    \targ       &    \qw  &\rstick{\raisebox{1.6em}{$\hspace{-.1cm}(\xgate^a  \zgate^{b +d} \otimes \xgate^{a + c} \zgate^{d})\cnot\ket{\psi}$}}
   \gategroup{1}{3}{2}{3}{1.5em}{\}}
    \gategroup{1}{1}{2}{1}{1.5em}{\{}
} } \caption{\label{fig:CNOT-gate}Protocol for a \cnot-gate. Here,
$\ket{\psi}$ is a two-qubit system.}
\end{figure}

\subsection{Protocol for a non-Clifford group gate}
\label{sec:non-clifford}

The only remaining gate required to implement universal quantum
computation is a non-Clifford group gate. We choose the
$\rgate$-gate.

\begin{figure}[ht]
\centerline{ \Qcircuit @C=2em @R=1.5em {
   \lstick{\xgate^a \zgate^b\ket{\psi}}   &   \gate{\rgate}    &       \rstick{\xgate^a \zgate^{a \oplus b} \pgate^a \rgate \ket{\psi }} \qw
} } \caption{\label{fig:R-gate-naive} First attempt at a protocol
for an \rgate-gate; output requires a $\pgate$ correction if $a=1$.}
\end{figure}

Our first attempt at a protocol for an \rgate-gate
(Figure~\ref{fig:R-gate-naive}) follows the protocols given in the
previous section: the server simply applies the \rgate-gate to the
encrypted data. However, this does not immediately work, since
$\rgate\xgate= \xgate \zgate\pgate\rgate$ and so in the case that an
$\xgate$-encryption is present, the output picks up an undesirable
$\pgate$ gate (this cannot be corrected by applying Pauli
corrections). In~\cite{C05}, Childs arrives at this same conclusion,
and then makes the astute observation that, in the case where $a=1$,
the server could be made to \emph{correct} this erroneous
$\pgate$-gate by executing a correction (which consists of
$\zgate\pgate$). As long as the server does not find out if this
correction is being executed or not, security holds.

\begin{figure}[t]
\centerline{
 \Qcircuit @C=1em @R=1em  {
&&&&&&\lstick{\xgate^a \zgate^b\ket{\psi}} & \qw & \gate{\rgate} & \qw & \targ &\qw    &  \meter & \control \cw \cwx[4]  &    &   \\
\lstick{\text{Server}}&&& & &&& & \control  & \cw    & \cw
&   \cw & \control \cw  & &&\\
&&&&& & & && & \ctrl{-2} &\qw    & \gate{\pgate^x} \cwx &  \qw  &
\qw& \rstick{\xgate^{a \oplus c} \zgate^{a(c\oplus y \oplus 1) \oplus b \oplus d \oplus y} \rgate \ket{\psi}}\\
&&&&\dw&\dw& \dw & \dw & \dw &\dw & \dw
 &  \dw & \dw  & \dw &\dw&\\
\lstick{\raisebox{-1cm}{\text{Client}}}&&& && &\lstick{x=a\oplus y}
&\cw & \control \cw \cwx[-3] & &
&   & & \control&  \cw& \rstick{c} \\
&&&& &&\lstick{\ket{+}}    & \gate{\pgate^y} & \gate{\zgate^d} & \qw
\qwx[-3]
 &   & & &\mbox{($y, d \in_R \{0,1\}$)} & &&\rstick{}
 \gategroup{1}{2}{3}{2}{0.7em}{\{}
 \gategroup{5}{2}{6}{2}{0.7em}{\{}
 }
 }
 \caption{\label{fig:R-gate}
Protocol for an $\rgate$-gate.
}
\end{figure}

This is where our approach takes a significantly different route
compared to~\cite{C05} or even~\cite{DNS10}: while these references
solve this problem
with \emph{two-way} quantum communication, 
we solve it with classical interaction and a single forward
auxiliary qubit randomly chosen out of four~possibilities. See
Figure~\ref{fig:R-gate}, as well as the proof of correctness in
 Appendix~\ref{sec:appendix-R-gate}.

Compared to~\cite{BFK09}, we manage to halve the size of the set
from which random qubits are chosen (from eight to~four). This can
be seen as due to the fact that~\cite{BFK09} directly implements a
hidden~$\rgate$-gate, while in Figure~\ref{fig:R-gate}, the server
first applies the $\rgate$-gate, and applies a correction by
performing a hidden $\pgate$-gate, requiring less resources than a
hidden $\rgate$-gate.
%

\subsection{Correctness and Security} \label{sec:correct-secure}
Given that the correctness of each gadget has been shown,
correctness of the main protocol is obvious: after each gadget, the
client adapts her knowledge of the keys used to encrypt the system
according to Figure~\ref{fig:aux-prep}--\ref{fig:R-gate}; each
gadget itself is correct, so the entire protocol implements the
quantum circuit as desired.

Our protocol provides the same level of security as the one-time pad
(Section~\ref{sec:preliminaries}), that is, it provides perfect
(information-theoretic) privacy. 
The rest of this section formalizes the definition of privacy based
on simulations and gives a proof based on the technique of giving an
equivalent, entanglement-based protocol (see
Section~\ref{sec:contributions}). For our definition of privacy, we
have used notions similar to those introduced by Watrous in the
context of quantum zero-knowledge interactive proof
systems~\cite{W09}.

Formally, a protocol for delegated computation is specified by a
pair $(C,S)$ representing an honest client and an honest server
(without loss of generality, both parties are quantum). As the
client is always honest, the security property concerns interactions
between pairs $(C, S')$ where $S'$ deviates arbitrarily from~$S$. At
the onset of the protocol, both parties agree on the classical
input~$q$ which determines the general quantum circuit to be
executed as an ordered series of gates acting on specified wires.
The structure of the interaction between $C$ and $S$ is thus
determined by~$q$. At the same time, a quantum input $\rhoin \in
\dens{\hreg{C}\otimes\hreg{S}}$ is distributed, $C$ receiving the
register in $\hreg{C}$ and $S$ receiving the register in $\hreg{S}$.
A cheating server $S'$ is any quantum computational process that
interacts with $C$ according to the message structure determined
by~$q$. By allowing $S'$ access to the input register~$\hreg{S}$, we
explicitly allow $S'$ to share prior entanglement with  $C$'s input;
this also models any \emph{prior} knowledge of~$S'$ and formalizes
the notion that the protocol cannot be used to \emph{increase}
knowledge.

Let $\hreg{Z}$ denote the output space of $S'$ and let
\added{$\Phi_q :
\linop{\hreg{S} \otimes \hreg{C}} \rightarrow \linop{\hreg{Z}}$}
be the mapping
induced by the interaction of $S'$ with $C$. Security is defined in
terms of the existence of a \emph{simulator} $\simul{S'}$ for a
given server $S'$, which is a general quantum circuit that agrees
with~$S'$ on the input and output dimensions. Such a simulator does
not interact with~$C$, but simply induces a mapping
\added{$\Psi_q : \linop{\hreg{S} \otimes \hreg{C}} \rightarrow \linop{\hreg{Z}}$ given by $\simul{S'}(\text{tr}_{\hreg{C}} \rhoin)$} on each input~$q$.
Informally, $(C,S)$ is private if the two mappings, $\Phi_q$ and
$\Psi_q$ are indistinguishable for every choice of~$q$ and every
choice of~$\rhoin$. Allowing for an $\epsilon$ amount of leakage, we
formalize this as the following:
A protocol $(C,S)$ for a delegated quantum
computation is \emph{$\epsilon$-private} if for every server~$S'$ there
exists a simulator $\simul{S'}$ such that for every classical
input~$q$,$\diam{\Phi_q- \Psi_q} \leq \epsilon\,,$
where~$\Phi_q$ is the mapping induced by the interaction of $S'$
with the client~$C$ on input~$q$ and $\Psi_q$ is the mapping induced
by $\simul{S'}$ on input~$q$. \looseness=-1

Taking $\epsilon=0$ gives the strongest possible security against a
malicious server: it does not allow for even an $\epsilon$ amount of
leakage, and allows the server to deviate arbitrarily (without
imposing any computational bounds). This is the level of security that we claim for our protocol for delegated quantum computation:
it is
$\epsilon$-private, with $\epsilon=0$. The proof follows \added{(although we do not formalize this notion here, we note that our proof method provides a simulator with the often-desirable property that it runs
with essentially the same computational resources as the deviating server).}

Fix a value for~$q$. We construct a simulator $\simul{S'}$ by giving
instructions how to prepare messages that replace the messages that
the client $C$ would  send to the server~$S'$ in the real protocol.
Privacy follows since we will show that these transmissions are
identical to those in the real protocol.

A high-level sketch of the proof is that we modify the behaviour of
the client in the main protocol (\textbf{Protocol 1}) in
a way that the effect of the protocol is unchanged \added{(meaning that both the output of the protocol and the view of the server is unchanged)}, yet the client
delays introducing her input into the protocol until after her
interaction with the server has ended (this makes the simulation
almost trivial). In order to do so, we describe below an
entanglement-based protocol (\textbf{Protocol~2}) as well as a
delayed-measurement protocol (\textbf{Protocol~3}).

We first consider \textbf{Protocol~2}, which is an
entanglement-based version of \textbf{Protocol~1}. In
\textbf{Protocol~2},  we modify how the client prepares her
messages, without modifying the server's actions or the effect of
the protocol. Thus, the preparing and sending of an encrypted
quantum register in step~\ref{prot:1} of
\textbf{Protocol~\ref{prot:1}}
  is replaced by an equivalent
teleportation-based protocol, as given in
Figure~\ref{fig:encrypt-teleport}. Also, the   $\rgate$-gate
protocol in step~\ref{prot:2} of \textbf{Protocol~1}
is replaced by an equivalent protocol as given in
Figure~\ref{fig:R-gate-EPR}. The protocol of
Figure~\ref{fig:R-gate-EPR} can be seen to be correct via an
intermediate protocol (Figure~\ref{fig:R-gate-Inter}), in which the
classical bit $x$ from the client to the server becomes a uniformly
random bit; this transformation is possible because in the protocol
of Figure~\ref{fig:R-gate}, $x = a \oplus y$ with $y$ a random bit.
Then choosing $x$ to be random and $y= a \oplus x$ gives an
equivalent protocol. The final entanglement-based protocol of
Figure~\ref{fig:R-gate-EPR} is seen to be correct via the circuit
identity given in Figure~\ref{fig:qubit-prep}. The remaining
protocols for stabilizer circuit elements are non-interactive and
thus unchanged in \textbf{Protocol~2}.

The main advantage of considering \textbf{Protocol~2} instead of
\textbf{Protocol~1} is that we can delay all the
client's measurements (in Figures~\ref{fig:encrypt-teleport}
and~\ref{fig:R-gate-EPR}) until the output register is returned in
step~\ref{prot:3} of \textbf{Protocol~1}, without
affecting the computation or the server's view of the protocol
(because actions on different subsystems commute); call the result
\textbf{Protocol~3}. In this delayed-measurement protocol, the
messages from the client to the server can be chosen \emph{before}
any interaction with the server, and are thus clearly independent of
the actions of~$S'$.

Thus we construct a simulator $\simul{S'}$ that plays the role of
the client in \textbf{Protocol~3}, \emph{but that never performs any
measurements} (thus, access to the actual input is not required). By
the argument above, $\simul{S'}$ actually prepares the same
transmissions as would $C$ in \textbf{Protocol~1}
interacting with $S'$ on any input~$\rhoin$.
It follows that simulating $S'$ on
these transmissions will induce the same mapping as $S'$ in the real
protocol, and thus $\diam{\Phi_q - \Psi_q} =0$\,. 

\begin{figure}[t!]
\centerline{
 \Qcircuit @C=.7em @R=1em  {
\lstick{\text{Server}}&&& & &&&&& &  \qw  & \qw &  \qw  & \qw &\rstick{\xgate^a\zgate^b\ket{\psi}} \\
&&&&\dw&\dw& \dw & \dw & \dw &\dw & \dw
 &  \dw & \dw  & \dw &\dw&\\
\lstick{\raisebox{-1cm}{\text{Client}}}&&& &&
&\lstick{\ket{0}} & \gate{\hgate} &\ctrl{1}      & \qw \qwx[-2] &      \\
&&&& &&\lstick{\ket{0}}    & \qw &\targ  &   \qw  &
\targ & \qw& \meter & \cw   &  \rstick{a} \\
&&&& &&\lstick{\ket{\psi}}    & \qw &\qw &   \qw  & \ctrl{-1} &
\gate{\hgate} & \meter & \cw & \rstick{b}
  \gategroup{3}{5}{4}{9}{1.4em}{--}
 \gategroup{3}{2}{5}{2}{0.7em}{\{}
 }
 }
 \caption{\label{fig:encrypt-teleport}Protocol to encrypt and send a qubit using teleportation\cite{BBCJPW93}.  The circuit in the dashed box prepares an EPR-pair. }
\end{figure}
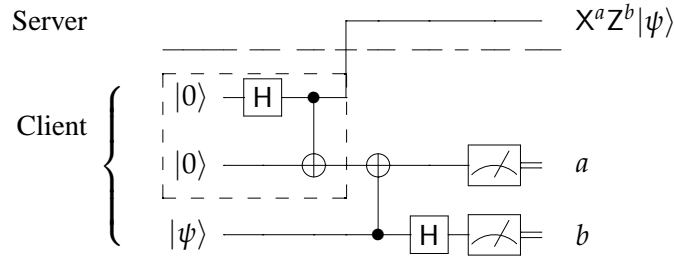

\begin{figure}[t!]
\centerline{
 \Qcircuit @C=1em @R=1em  {
&&&&&&&\lstick{\xgate^a \zgate^b\ket{\psi}} & \qw & \gate{\rgate} & \qw & \targ &\qw    &  \meter & \control \cw \cwx[4]  &    &   \\
\lstick{\text{Server}}&&& & && &&& \control  & \cw    & \cw
&   \cw & \control \cw  & &&\\
&&&& & & && &&& \ctrl{-2} &\qw    & \gate{\pgate^x} \cwx &  \qw  &
\qw& \rstick{\xgate^{a \oplus c} \zgate^{a(c\oplus y \oplus 1) \oplus b \oplus d \oplus y} \rgate \ket{\psi}}\\
&&&\dw&\dw&\dw&\dw& \dw & \dw & \dw &\dw & \dw
 &  \dw & \dw  & \dw &\dw&\\
\lstick{\raisebox{-1cm}{\text{Client}}}&& &&&& &\lstick{x \in_R
\{0,1\}} &\cw & \control \cw \cwx[-3] & &
&   & & \control &  \cw& \rstick{c} \\
&&&&& &&\lstick{\ket{+}}    & \gate{\pgate^y} & \gate{\zgate^d} &
\qw \qwx[-3]
 &   & & &\mbox{($d \in_R \{0,1\}, y = a \oplus x $)} & &&
 \rstick{}
 \gategroup{1}{2}{3}{2}{0.7em}{\{}
 \gategroup{5}{2}{6}{2}{0.7em}{\{}
 }
 }
 \caption{\label{fig:R-gate-Inter}Intermediate Protocol for an $\rgate$-gate. Compared to
Figure~\ref{fig:R-gate}, the classical message from the client to
the server is chosen uniformly at random. This protocol performs the
same computation as the protocol in Figure~\ref{fig:R-gate}.}
\end{figure}
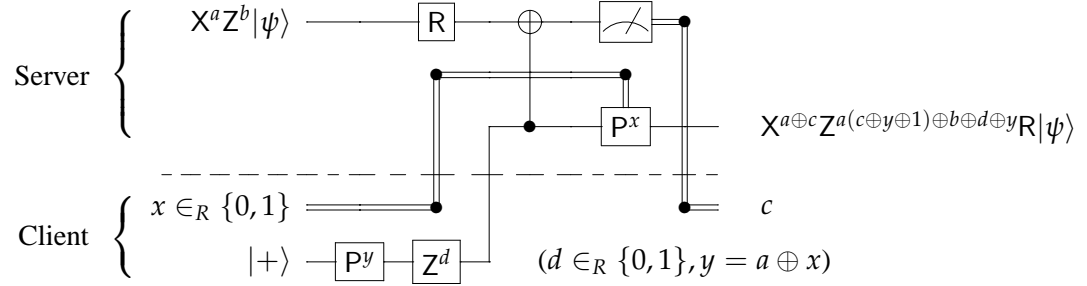

\begin{figure}[t!]
\centerline{
 \Qcircuit @C=1em @R=1em  {
&&&&&&&\lstick{\xgate^a \zgate^b\ket{\psi}} & \qw & \gate{\rgate} & \qw & \targ &\qw    &  \meter & \control \cw \cwx[4]  &    &   \\
\lstick{\text{Server}}&&&&& & && & \control  & \cw    & \cw
&   \cw & \control \cw  & &&\\
&&&&&& & & && & \ctrl{-2} &\qw    & \gate{\pgate^x} \cwx &  \qw  &
\qw& \rstick{\xgate^{a \oplus c} \zgate^{a(c\oplus y \oplus 1) \oplus b \oplus d \oplus y} \rgate \ket{\psi}}\\
&&&&\dw &\dw&\dw& \dw & \dw & \dw &\dw & \dw
 &  \dw & \dw  & \dw &\dw&\\
&& &&&& &\lstick{x \in_R \{0,1\}} &\cw & \control \cw \cwx[-3] & &
&   & & \control &   \cw  & \rstick{c} \\
\lstick{\text{Client}}&&& &&&&\lstick{\ket{0}}    & \gate{H}
&\ctrl{1} & \qw \qwx[-3]
 &   &\mbox{($y = a \oplus x $)} & & & &&\\
&& &&&& &\lstick{\ket{0}}    & \qw & \targ & \qw
 & \gate{\pgate^y}  & \gate{\hgate} &  \meter &  \cw  & \cw
   & \rstick{d}
    \gategroup{6}{6}{7}{10}{1.4em}{--}
     \gategroup{1}{2}{3}{2}{0.7em}{\{}
 \gategroup{5}{2}{7}{2}{0.7em}{\{}
 }
 }
  \caption{\label{fig:R-gate-EPR}Entanglement-based protocol for an $\rgate$-gate. This protocol
performs the same computation as the protocols in
Figures~\ref{fig:R-gate} and~\ref{fig:R-gate-Inter}. The circuit in
the dashed box prepares an EPR-pair.}
\end{figure}
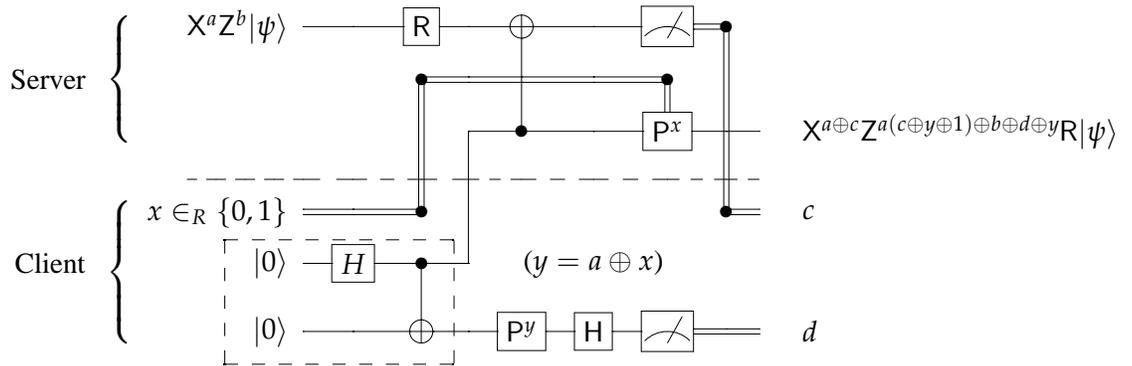


Note that a malicious server may not necessarily follow the
protocol, thus possibly interfering with the computation. Our
protocol does not guard against this; detecting a cheating server
can be done using a \emph{quantum authentication
code}~\cite{BCGST02}, as is done in~\cite{ABE10} (albeit by
requiring the client to have more quantum power). It is important to
note however that our privacy definition and proof holds against
such a malicious server.

\section*{Acknowledgements}
I would like to thank Richard Cleve, Serge Fehr, Stacey Jeffery and
Debbie Leung for related discussions. This work was completed while the author was affiliated with
the Canadian Institute for Advanced Research, the University of Waterloo and the Institute for
Quantum Computing.

\bibliographystyle{plain}

\appendix

\section{Correctness of the \rgate-gate protocol}
\label{sec:appendix-R-gate}
\renewcommand{\figurename}{Circuit}
\setcounter{figure}{0}

We give below a step-by-step proof of the correctness of the
\rgate-gate protocol as given in Figure~\ref{fig:R-gate}. The basic
building block is the circuit identity for an \xgate-teleportation
from~\cite{ZLC00}, which we re-derive here.

\begin{enumerate}
\item  Our first circuit identity swaps a qubit $\ket{\psi}$ with the
state $\ket{+}$ and is easy to verify.

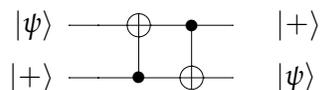
\begin{figure}[h!]
\centerline{
 \Qcircuit @C=1em @R=1em {
\lstick{\ket{\psi}} & \qw &\targ      &  \ctrl{1} & \qw   &  \rstick{\ket{+}}     \\
\lstick{\ket{+}}    & \qw &\ctrl{-1}  &   \targ   & \qw &
\rstick{\ket{\psi}}&   }
 }
 \label{circuit-proof:1}
\end{figure}

\item We can measure the top qubit in the above circuit
and classically control the output correction. We have thus
re-derived the circuit corresponding to the ``\xgate-teleportation''
of~\cite{ZLC00}.

\begin{figure}[h!]
 \centerline{
 \Qcircuit @C=1em @R=1em {
\lstick{\ket{\psi}} & \qw &\targ      &  \meter & \cw   &  \rstick{c}      \\
\lstick{\ket{+}}    & \qw &\ctrl{-1}  &   \qw  & \qw &
\rstick{\xgate^c\ket{\psi}}&   }
 }

\end{figure}

\item Next, we re-define the input to be $\rgate \xgate^a
\zgate^b\ket{\psi}$, so the output becomes $\xgate^c \rgate
\xgate^a\zgate^b \ket{\psi} = \xgate^{a \oplus c}\zgate^{a \oplus b}
\pgate^a \rgate \ket{\psi}$.

\begin{figure}[ht!]
\centerline{
 \Qcircuit @C=1em @R=1em {
\lstick{\xgate^a \zgate^b\ket{\psi}} & \gate{\rgate} &\targ      &  \meter & \cw   &  \rstick{c}      \\
\lstick{\ket{+}}    & \qw &\ctrl{-1}  &   \qw  & \qw &
\rstick{\xgate^{a \oplus c}\zgate^{a \oplus b} \pgate^a \rgate
\ket{\psi}}& }
 }
\end{figure}

%

\item Then add three gates ($\pgate^y$, $\zgate^d$, $\pgate^{a \oplus y}$) to the bottom wire (see circuit below). Applying identities from Section~\ref{sec:preliminaries}, we
get as output what we expect:
\begin{align*}
&\pgate^{a \oplus y} \zgate^d \pgate^y \xgate^{a \oplus
c}\zgate^{a\oplus b} \pgate^a \rgate \ket{\psi}\\
 &= \zgate^{a\cdot y}
\pgate^{a+y} \zgate^{d} \pgate^y \xgate^{a\oplus c}\zgate^{a\oplus
b} \pgate^a
\rgate \ket{\psi} \\
&= \zgate^{d \oplus a \cdot y \oplus y}  \pgate^{a} \xgate^{a\oplus
c}\zgate^{a\oplus b} \pgate^a
\rgate \ket{\psi} \\
&= \zgate^{d \oplus a \cdot y \oplus y} \xgate^{a \oplus
c}\zgate^{a(a \oplus c)} \pgate^a \zgate^{a\oplus b}
\pgate^a \rgate \ket{\psi} \\
&= \xgate^{a \oplus c} \zgate^{d \oplus a\cdot y \oplus y \oplus a^2
\oplus a\cdot c} \zgate^{b} \rgate \ket{\psi}\\
 &= \xgate^{a \oplus c} \zgate^{a(c\oplus y\oplus 1)
\oplus b \oplus d \oplus y} \rgate \ket{\psi}
\end{align*}

\begin{figure}[h!]
\centerline{
 \Qcircuit @C=1em @R=1em {
\lstick{\xgate^a \zgate^b\ket{\psi}} & \gate{\rgate} & \qw &\targ     &  \meter &  \cw  &  \rstick{c}      \\
\lstick{\ket{+}}    & \gate{\pgate^y} & \gate{\zgate^d} & \ctrl{-1}
 & \gate{\pgate^{a\oplus y}}  &\qw & \rstick{\xgate^{a \oplus c} \zgate^{a(c\oplus y \oplus 1) \oplus b \oplus d \oplus y} \rgate \ket{\psi}}& }
 }

\end{figure}

\end{enumerate}

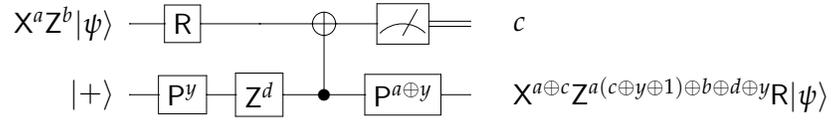
\end{document}